\documentclass[prb,nofootinbib,noshowkeys,twocolumn,superscriptaddress]{revtex4}
\usepackage[latin1]{inputenc}
\usepackage{pstricks,pst-node,pst-text,pst-3d,pslatex}
\usepackage[T1]{fontenc}
\usepackage{amsmath}

\usepackage{graphicx,amsfonts}
\usepackage{amsmath}

\newcommand{\be}{\begin{eqnarray}}
\newcommand{\ee}{\end{eqnarray}}

 \newcommand{\D}{\mathcal{D}}
\begin{document}
\author{L. Boninsegna}
\affiliation{Dipartimento di Fisica  Universit\'a degli Studi di Trento, Via Sommarive 14, Povo (Trento), I-38050 Italy}
\author{P.~Faccioli}
\email{faccioli@science.unitn.it}
\affiliation{Dipartimento di Fisica  Universit\'a degli Studi di Trento, Via Sommarive 14, Povo (Trento), I-38050 Italy}
\affiliation{ INFN, Gruppo Collegato di Trento, Via Sommarive 14, Povo (Trento), I-38050 Italy} 

\title{Quantum Charge Transport and Conformational Dynamics of Macromolecules} 

\begin{abstract}
We study the dynamics of quantum excitations inside macromolecules which can undergo  conformational transitions.
In the first part of the paper,  we use the path integral formalism to rigorously derive a set of coupled equations of motion which
 simultaneously describe the molecular and quantum transport dynamics,  and obey the  fluctuation/dissipation relationship.
 We also  introduce an algorithm  which yields the most probable molecular and quantum transport pathways in  rare, 
  thermally-activated  reactions. In the second part of the paper, we apply this formalism to simulate  the propagation of a quantum charge during the collapse of a polymer from an initial stretched conformation
 to a final globular state. 
We find that the charge dynamics is quenched when the chain reaches a molten globule state. Using random matrix theory we show that this transition is
 due to an increase of quantum localization driven by dynamical disorder.   
 \end{abstract}
\maketitle
\section{Introduction}
Understanding the mechanisms involved in the transport of charged and neutral quantum excitations inside macromolecules is a key step towards realizing nano-scale  organic devices with functional 
(opto-)electronical properties, notably molecular wires and antennas.  
 This perspective has motivated a huge activity devoted to investigating the conductance of
 inorganic\cite{inorganic1, inorganic2, inorganic3, inorganic4, inorganic5, inorganic6}, organic~\cite{organic1}  and biological \cite{DNAbook, DNAelsner, peptide1, DNAreview} polymers.  

Unlike quantum wires made of solid-state nano ribbons~\cite{nanoribbons1,nanoribbons2, nanoribbons3,nanoribbons4}, flexible molecular wires in solution  can undergo  
conformational transitions. This feature raises the problem of understanding the 
 implications of the conformational dynamics on the quantum transport properties of the molecule.  
Recent experimental and theoretical studies have shown that even small thermal fluctuations of the DNA backbone can significantly alter its conductivity~\cite{DNAbook, DNAelsner}.
These effects should be greatly amplified in more flexible polymers, which can undergo  cooperative transitions and major re-arrangements of the three-dimensional structure. 

Investigating the quantum charge transport dynamics during a coil-globule transition of a flexible chain provides an ideal framework to probe
 the conformation/conductance relationship and assess the role of quenched and dynamical disorder. In particular, studying such a process involves the exploration of the
  crossover from a regime in which the quantum transport dynamics is effectively one-dimensional  to one in which quantum excitations can diffuse in three dimensions. 
  In addition, understanding under which conditions a 
 molecular wire can display different conducting behavior in the swollen and collapsed phases may lead to designing nano-scale molecular switches, which can be activated by means of 
 chemically- or thermally- induced unfolding. 

From a theoretical and microscopic standpoint, the study of quantum transport in molecular systems requires a formalism  
in which the electronic excitations are explicitly taken into account and the atomic nuclei are coupled to a solvent.  
In the existing approaches~\cite{book, DNAbook, Wallace, FO}, the equations of motion which describe the time-evolution of the charge density and atomic coordinates are 
postulated phenomenologically. For example, in Ref. \cite{DNAelsner} 
the DNA conductivity was computed by assuming that the atomic nuclei obey Newton's equation with a classical inter-atomic force field. In addition, charged groups in the DNA backbone and counter ions in the solvent were coupled electrostatically with the quantum charge. 
Such an approach neglects all non-Coulombic interactions between quantum and classical degrees of freedom. For example, the molecule may lower the total energy by assuming configurations in which the ionization energy of the quantum charges is increased. This interaction has in principle implications on the molecular dynamics.  
 
In view of these considerations, it would be valuable if the  equations describing the dynamics of electronic excitations and nuclear coordinates were rigorously derived starting from the general theoretical framework of  open quantum systems~\cite{petruccione}.  On the one hand, this would guarantee that all the correlations between the quantum and classical degrees of freedom are
consistently taken into account.    On the other hand, it would ensure the fluctuation-dissipation relationship is respected, hence that the correct thermodynamics is recovered in the long-time limit. 

In the first part of this paper, we use the Feynman-Vernon path integral formalism to  provide such a derivation and develop a rigorous microscopic theory for the dynamics 
of quantum excitations in molecular systems  in solution.
The same path integral formalism is also used to derive an algorithm which efficiently yields the most probable 
molecular and quantum transport pathways in rare thermally-activated  reactions.  

In the second part of this paper, we use our quantum  and stochastic equations of motion to investigate the  
propagation of a quantum charge inside a homo-polymer undergoing a coil-globule transition.
We find that the charge dynamics is strongly suppressed due to a quantum localization driven by the dynamical disorder. Interestingly, this effect sets in only
when the
polymer reaches a compact conformation. 
 
The paper is organized as follows.  In section \ref{Hamiltonian} we define define the quantum  Hamiltonian for system. 
 In section, \ref{FV} we construct the 
path integral representation  of the time-dependent probability to simultaneously observe  the molecule at  a given conformation and the charge at a given molecular site. This 
path integral is then analyzed in section \ref{Saddlepoint}, in saddle-point approximation, and in section \ref{QMD} 
we derive the set of  equations of motion for the  system and an algorithm to predict the most probable pathways in thermally activated transitions. 
Section \ref{application} is devoted to the study of the collapse of the chain. Results, conclusions and outlooks are summarized in  section \ref{conclusions}. 

\section{A microscopic model for quantum transport in  macromolecules}
\label{Hamiltonian}

Quantum transport processes  in dynamical molecular systems 
depend on the time-dependent structure of the electronic ground-state and excited states. 
Clearly, an  approach in which these wave-functions and the nuclear coordinates are self-consistently calculated  \emph{ab-initio} would be 
extremely computationally expensive.  

A commonly efficient strategy to reduce this computational complexity consists in
coarse-graining of the electronic problem into an effective configuration-dependent tight-binding Hamiltonian.  
In particular, in the so-called Fragment Orbital  approach \cite{FO}, one first identifies a set of molecular sites (so-called fragments) where the charge (i.e. a  hole) can be found.  
For example,  in DNA these sites can be identified with with individual Watson-Crick bases or with base-pairs. 
Then, the highest occupied molecular orbitals  for each  fragment --- herby denoted with $\phi_i$ ($i=1, \ldots N_s$)--- 
are computed in density functional theory and Born-Oppenheimer approximation, neglecting the effect of the rest of the molecule.

The  hopping of the charge between different molecular fragments is controlled by the standard tight-binding Hamiltonian:
\be
\label{HMC}
\hat H_{MC} = \sum_{l,m=1}^{N_s} f_{lm}~\hat a^\dagger_l \hat a_m,
\ee 
where $f_{l m} \equiv T_{l m} - e_l\delta_{lm}$. The parameters  $T_{l m}$ and $e_l$ are obtained from the fragment orbitals  $|\phi_l\rangle$ and 
$|\phi_m\rangle$:  
\be\label{TFO1}
 T_{l m} &\equiv& \langle \phi_l |\hat H_{KS} |\phi_m \rangle, \\
 \label{TFO2}
 e_{l} &\equiv& \langle \phi_l | \hat H_{KS} |\phi_l \rangle,
 \ee 
where $\hat H_{KS}$ is the Kohn-Shawn Hamiltonian.  The hopping parameters $T_{lm}$ and the on-site energies $e_l$ can be taken to be real-valued. 

In addition, if the molecule is connected to electrodes, the Hamiltonian $\hat H_{MC}$  must include also the coupling with the donor and the acceptor:
\be
\hat H_{MC} &\rightarrow& \hat H_{MC} + (T_{D 1} \hat a^\dagger_{1} \hat a_{D} + h.c.) +(T_{N_s A} \hat a^\dagger_{A} a_{N_s} + h. c) \nonumber\\
&&- e_D   \hat a^\dagger_{D} \hat a_{D}  - e_A   \hat a^\dagger_{A} \hat a_{A}
\ee

In the Born-Oppenheimer approximation, the fragment orbitals in Eq.s (\ref{TFO1}) and (\ref{TFO2})
depend parametrically on the coordinates of the atomic nuclei, which evolve in time under the effect of
the inter-atomic forces,  and of the interactions with quantum charge and with the solvent. 

A consistent way to describe this dynamics is to consider the fully quantum Hamiltonian:
\be
\label{Htot}
\hat H = \hat H_{MC} + \hat H_M + \hat H_B + \hat H_{MB}.
\ee
In this equation,   $\hat H_{MC} $ is the tight-binding Hamiltonian defined in Eq. (\ref{HMC}), while the Hamiltonian $\hat H_M$ controls the conformational dynamics of the molecule
in the absence of quantum excitations, 
\be
\label{HM}
\hat H_M \equiv \sum_{\alpha=1}^{N_p} \frac{\hat p_\alpha^2}{2 M} + \hat V(Q),
\ee
 where $Q=(q_1, \ldots, q_{N_p})$ is the set of all $N_p$ atomic coordinates  and $V(Q)$ is the molecular 
 potential energy which includes the interaction between the different atoms and possibly a  term to account for the electrostatic and hydro-phobic/philic 
 interaction with the solvent.  

The part of the Hamiltonian $\hat H_B + \hat H_{MB}$ describes
the coupling of the molecule with a thermal heat-bath,  modeled with an infinite set of harmonic-oscillators: 
\be
\hat H_{B} &=& \sum_{\alpha=1}^{N_p}\sum_{j=1}^{\infty}\left(\frac{\hat \pi_j^2}{2\mu_j}+\frac{1}{2}\mu_j \omega_j^2 \hat x_j^2  \right)\\
\label{HMB}
\hat H_{MB} &=& \sum_{\alpha=1}^{N_p}\sum_{j=1}^{\infty}\left(- c_j \hat x_j \hat q_\alpha + 
~ \frac{c_j^2}{2\mu_j \omega_j^2}\hat q_\alpha^2\right).
\ee
$X=(x_1, x_2, \ldots)$ and $\Pi = (\pi_1, \pi_2, \ldots)$ are the harmonic oscillator coordinates and momenta, 
 $\mu_j$ and $\omega_j$ denote their masses and 
 frequencies and $c_j$  are the couplings between atomic and heat-bath variables.
  The last term in Eq. (\ref{HMB})  is a  standard counter-term introduced to compensate the renormalization of the molecular potential energy which occurs when the heat-bath variables
  are traced out (see e.g. discussion in Ref. \cite{PIreview}).   
  
 The Hamiltonian (\ref{Htot}) describes a close system at the  fully quantum level. In the next sections, we shall use the path integral formalism to trace out the heat-bath variables
  and take the classical limit for the nuclear degrees of freedom.
 
  \section{Path integral representation of the quantum-diffusive dynamics of the system}
\label{FV}

In this section we derive a the path integral representation of the time-evolution of the system described by the Hamiltonian~(\ref{Htot}).
  
Let us assume that the molecule is  prepared in some configuration $Q_{0}$ and that  a hole  is initially created at 
some monomer site $k_{0}$. We are interested in computing the conditional probability  $P_t(k_f,Q_f| k_0, Q_0)$ that after a time interval $t$ the molecule is found in  
conformation $Q_f$ and  the charge at the site $k_f$.  Such a probability is described by the following time-dependent reduced quantum density matrix: 
\be
\label{Pcond}
P_t(k_f,Q_f,| k_0, Q_0) &=& \frac{\text{Tr} [|k_f, Q_f\rangle \langle k_f Q_f| \hat \rho(t)]}{\text{Tr} ~\hat \rho(t)}\nonumber\\
&=&
\frac{\text{Tr} [|k_f, Q_f\rangle \langle Q_f, k_f| 
 e^{-\frac{i}{\hbar}\hat H t}~\hat \rho(0)~e^{\frac{i}{\hbar}\hat H t}]}
 {\text{Tr}~ \hat \rho(0)},\qquad
 \ee
where  $\hat \rho(0) = | Q_0 k_0\rangle \langle Q_0 k_0 | ~e^{-\frac{1}{K_B T}\hat H_{B}}$ is the initial density matrix, which assumes factorization with a
thermal distribution for the heat-bath variables.   

The goal of this section is to represent the conditional probability (\ref{Pcond}) as a path integral. This can be done relying on the so-called Feynman-Vernon  
formalism for open quantum systems, which has been extensively  applied to study quantum brownian motion~\cite{PIreview}. 

The path integral representation of the density matrix can be constructed in a way which is conceptually similar to that used to represent the standard Feynman
propagator. The main difference is that it requires to perform
the  Trotter decomposition of two real-time evolution operators (forward and backwards in time) and an one imaginary time evolution (initial thermal distribution of the heat-bath 
variables). 

It is convenient to adopt a field-theoretic language to represent the dynamics of the quantum charge, while using the position representation to describe the evolution of the
 atoms in the molecule and of the harmonic oscillators in the heat-bath. In practice, this corresponds to introducing the following resolution of the identity at each slide in the Trotter decomposition 
of the real- and imaginary- time evolution operators entering Eq. (\ref{Pcond}):
\be
1 &=& \int d Q \int d X \int \left( \prod_{k=1}^{N_s} \frac{d\phi_k d\phi_k^*}{2 \pi i}\right)
e^{-\sum_{l=1}^{N_s} \phi_l \phi^*_l} |Q, X, \Phi \rangle,\nonumber\\
 \ee
where  $\Phi= (\phi_1, \ldots \phi_{N_s}) $  are the eigenvalues
 of the bosonic coherent states constructed from creation and annihilation operators in Eq. ~(\ref{HMC}). 
 
An advantage of adopting a field-theoretic representation for the quantum charge dynamics is that the statistical weight of the coherent field configurations in the path 
integral  can be written as the exponent of an action, i.e. 
$\int \D\phi \D \phi^* e^{\frac{i}{\hbar}S[\phi^*, \phi]}$. This property is very useful to develop the saddle-point approximation, and is not satisfied if the dynamics
of a tight-binding Hamiltonian is represented using the discrete position eigenstates. 

The conditional probability (\ref{Pcond}) in path integral form reads:
\be
\label{PI1}
&&P_t(k_f, Q_f | k_0, Q_0) = \int d\bar X ~\int dX_1\int dX_2 \int_{X_1}^{X_2} \D\tilde X e^{- S_E[\tilde X]}\nonumber\\
&& \int_{X_1}^{\bar X} \D X' \int_{X_2}^{\bar X} \D X''  \int_{Q_0}^{Q_f}\D Q'  \int \D \phi^{'} \D \phi^{'*} ~\phi^{'}_{k_f}(t)\phi^{' *}_{k_0}(0)
\nonumber\\
&& e^{-\sum_m \phi^{' *}_m(0)\phi^{'}_m(0)}~e^{\frac{i}{\hbar} \left(  S_{MC}[Q^{'},\phi^{'},\phi^{'*}] + S_{MB}[Q',X']\right)} \nonumber\\
&&\int_{Q_0}^{Q_f} \D Q'' \int\D \phi'' \D \phi^{''*} \phi^{'' *}_{k_f}(t)\phi^{''}_{k_0}(0) 
 \nonumber\\
&& ~e^{-\sum_m\phi^{''*}_m(t)\phi^{''}_m(t)}~ e^{-\frac{i}{\hbar}\left(S_{MC}[Q^{''},\phi^{''},\phi^{''*}] + S_{MB}[Q^{''},X^{''}]\right)},  
\ee 
In this equation $\phi^{'}_l(t), \phi^{''}_l(t)$ are the complex bosonic fields associated to the  charge coherent states propagating forward and backwards in time 
respectively and 
the action functionals appearing at the exponents read
\be
\label{SMC}
S_{MC}\left[Q, \phi, \phi^{*}\right]  &=& \int_0^t dt'\left[\frac{M \dot Q(t')^2}{2}-V[Q(t')]\right. \nonumber\\
&+&\left. \sum_{l,m}  \phi_l^*(t')\left( i\hbar\frac{\partial }{\partial t'}\delta_{l m} -f_{l m}[Q(t')] \right) \phi_m(t'). \right.\nonumber\\
 \\
S_{MB}[X, Q] &=& \int_0^t dt' \sum_j \left(\frac{\mu_j\dot{x}_j(t')^2}{2}-\frac{1}{2}\mu_j\omega_j^2 x_j(t')^2 \right)\nonumber\\
&+& \sum_\alpha \sum_j \left(c_j x_j(t')q_\alpha(t')-\frac{c_j^2}{2\mu_j\omega_j^2} q_\alpha(t')^2]\right),\nonumber\\ \\
\label{SE}
S_{E}[X] &=& \int_0^\beta d\tau \left[\sum_j \left( \frac{\mu_j\dot{x}_j(\tau)^2}{2}+\frac{1}{2}\mu_j\omega_j^2 x^2_j(\tau)\right)\right].\nonumber\\ 
\ee

The path integrals over the harmonic oscillator variables are Gaussian and can be carried out analytically. One obtains 
\be
\label{PI2}
&&P_t(k_f, Q_f | k_0, Q_0) =
  \int_{Q_0}^{Q_f}\D Q^{'}  \int \D \phi^{'} \D \phi^{'*} ~\phi^{'}_{k_f}(t)\phi^{' *}_{k_0}(0) \nonumber\\
&&e^{-\sum_m \phi^{' *}_m(0)\phi^{'}_m(0)+\frac{i}{\hbar}S[Q^{'},\phi^{'},\phi^{'*}]} \int_{Q_0}^{Q_f} \D Q'' \int\D \phi'' \D \phi^{''*}  
 \nonumber\\
&& \cdot \phi^{'' *}_{k_f}(0)\phi^{''}_{k_0}(t) \cdot ~e^{-\sum_m\phi^{''*}_m(t)\phi^{''}_m(t)-\frac{i}{\hbar}S_{MC}[Q^{''},\phi^{''},\phi^{''*}]} ~e^{-\Phi_{FV}[Q',Q'']}.\nonumber\\
\ee 
$\Phi_{FV}[Q',Q]$ is the so-called Feynman-Vernon influence functional\cite{PIreview}, which describes the fluctuation and dissipation 
induced by the coupling with the heat-bath and reads:  
\be
&&\Phi_{FV}[Q',Q'']=\frac{1}{\hbar}\int_0^t dt' \int_0^{t'} dt''\left[Q'(t')-Q''(t') \right] \nonumber\\
&&\left[L(t'-t'')Q'(t'') -L^*(t'-t'')Q''(t'') \right]\nonumber\\
&& + i\frac{\bar\mu}{2\hbar}\int_0^t dt'\left[{Q'}^2(t')-{Q''}^2(t') \right], \qquad  \left(\bar\mu = \sum_j \frac{c_j^2}{m_j \omega_j^2}\right).\nonumber\\
\ee
$L(t)$ is a two-point correlation function which encodes the physics of the coupling of the molecular coordinates with the heat-bath and reads:
\be
L(t) = \sum_j \frac{c_j^2}{\mu_j \omega_j}\left[\text{coth}\left(\frac{\omega_k \hbar}{2 k_B T}\right)~\text{cos}(\omega_j t)- i ~\text{sin}(\omega_j t)\right].
\ee
Note that the strength of the fluctuation and dissipation induced by the solvent and the time scales at which memory effects die out can be tuned by changing the
 parameters in the harmonic bath Hamiltonian (\ref{HMB}). In particular, here we consider the so-called ohmic bath limit ( see e.g. Ref. \cite{PIreview}),
 in which the  $L(t)$ reduces to 
 \be
 \label{Lohm}
L(t) \rightarrow L^{ohm}(t) =\frac{2 k_B T M \gamma}{\hbar} \delta(t)+ \frac{ i~ M \gamma}{2}~ \frac{d}{dt}\delta(t),
 \ee
 and $\gamma$ is interpreted as the friction coefficient.  In section \ref{QMD} we shall show that this choice leads to the natural generalization of the classical over-damped Langevin 
 dynamics.

\section{Saddle-point Approximation of the Path Integral} 
\label{Saddlepoint}

The path integral (\ref{PI2}) provides an exact representation of the conditional probability given in Eq. (\ref{Pcond}) and cannot be solved without relying on some approximation.  
To this end, we observe that for any time $t$  there exists an obvious sum rule:
\be\label{sumrule}
\sum_{k_f}\int d Q_f ~P_t(k_f, Q_f| k_0, Q_0) = 1.
\ee
The idea is then to use the path integral representation (\ref{PI2}) to implement this condition, and analyze it in saddle-point approximation.

We begin by changing variables for the molecular coordinates:
\be
y(t) \equiv Q^{'}(t)- Q^{''}(t)\quad r(t) = \frac{1}{2} (Q^{'}(t)+ Q^{''}(t)) 
\ee
Next, we use the  path integral (\ref{PI2}) to re-write the sum-rule (\ref{sumrule})  as
\be\label{PI3}
&&1  =
\sum_{k_f} \int_{Q_0} \D r \int_0^0 dy   \int \D \phi^{'} \D \phi^{'*} ~\phi^{'}_{k}(t)\phi^{' *}_{k_0}(0) \nonumber\\
&&e^{-\sum_m \phi^{' *}_m(0)\phi^{'}_m(0)+\frac{i}{\hbar}S_{MC}[r+y/2,\phi^{'},\phi^{'*}]}  \int\D \phi'' \D \phi^{''*}  
 \nonumber\\
&& \cdot \phi^{'' *}_{k_f}(0)\phi^{''}_{k_0}(t) \cdot ~e^{-\sum_m\phi^{''*}_m(t)\phi^{''}_m(t)-\frac{i}{\hbar}S_{MC}[r-y/2,\phi^{''},\phi^{''*}]} ~e^{-\Phi^{'}_{FV}[r,y]}\nonumber\\
\ee
where the new expression for the Feynman-Vernon functional reads
\be
\Phi^{'}_{FV}[r,y] =  \int_0^t dt'  \left(\frac{ M\gamma K_B T}{\hbar^2} y^2(t') +  \frac{ i M \gamma}{\hbar}~ \dot r \cdot y\right). 
\ee

We now introduce a set of tensor fields  $\rho^{'}_{l m}(t)$,$\rho^{''}_{l m}(t)$, $\sigma^{'}_{l m}(t)$ and $\sigma^{''}_{l m}(t)$ 
into the path integral  by means the functional identities 
\be
&&1 = \int \D \sigma^{'} \D\rho^{'}  e^{\frac{i}{\hbar} \sum_{l, m} \int_0^t dt'  \sigma^{'}_{l m}
 \left( \rho'_{l m} -\phi_l^{*'}\phi^{'}_m~\right)},
\nonumber\\
&&1 = \int \D \sigma^{''} \D\rho^{''}  e^{-\frac{i}{\hbar} \sum_{l,m} \int_0^t dt' \sigma^{''}_{l m} \left( \rho^{''}_{l m}-\phi_l^{*''}\phi^{''}_m~\right)}.
\nonumber\\
\ee

The new expression for the sum-rule  (\ref{sumrule}) is then
\be
\label{PI4}
&& 1 = \int \mathcal{D}\rho^{'} \mathcal{D}\rho^{''} \mathcal{D}\sigma^{'} \mathcal{D}\sigma^{'} 
 e^{\frac{i}{\hbar}\int_0^t dt' \sum_{l,m} (\sigma^{'}_{l m} \rho^{'}_{l m} - \sigma^{'}_{l m} \rho^{''}_{lm} )}\quad
 \nonumber\\
&& \int_{Q_0}  \mathcal{D}r \int \mathcal{D}y~e^{\frac{i}{\hbar}\mathcal{W}[r,y, \rho^{'}, \rho^{''}]}
e^{-\Phi^{'}_{FV}[r,y]} ~\left( \sum_{k_f} \mathcal{Q}_{k_f}[\sigma^{'}] ~ \mathcal{M}_{k_f} [\sigma^{''}]\right),\quad
\ee
where the functionals $Q_{k_f}$ and $M_{k_f}$ read
\be
\mathcal{Q}_{k_f}[\sigma^{'}] &=& \int \mathcal{D}\phi^{'} \mathcal{D}\phi^{'*} \phi^{'}_{k_f}(t) \phi^{'*}_{k_0}(0) e^{-\sum_m \phi^{'*}_m(0)\phi^{'}_m(0)} \nonumber\\
&& e^{\frac{i}{\hbar} S_{MF}[\phi^{'}, \phi^{'*}, \sigma^{'}]},\\
 \mathcal{M}_{k_f}[\sigma^{''}] &=& \int \mathcal{D}\phi^{''} \mathcal{D}\phi^{''*} \phi^{''*}_{k_f}(0) \phi^{''}_{k_0}(t) e^{-\sum_m \phi^{''*}_m(t)\phi^{''}_m(t)} \nonumber\\
&& e^{-\frac{i}{\hbar} S_{MF}[\phi^{''}, \phi^{''*}, \sigma^{''}]},
\ee
while the functional $\mathcal{W}$ and $S_{CM}$ are defined as
\be
\mathcal{W}[x, y, \rho^{'},\rho^{''}] &=& \int_0^t dt' \left\{M~ \dot r \dot y - V\left(r + \frac{y}{2}\right) +  V\left(r - \frac{y}{2}\right) \right.\nonumber\\
&-&\left. \sum_{l,m}\left[f_{l m}\left(r+\frac{y}{2}\right)~\rho^{'}_{lm}-f_{lm}\left(r-\frac{y}{2}\right)~\rho^{''}_{lm} \right] \right\},\nonumber\\ \\
S_{MF}[\phi,\phi^{*}, \sigma]&=& \sum_{l,m} \int_0^t dt'  \phi_l^*(t') \left[ i\hbar~ \frac{\partial}{\partial t'}\delta_{lm}
- \sigma_{lm}(t') \right]\phi_m(t').\nonumber\\
 \ee

The  path integral  (\ref{PI4}) is still and exact representation of the sum-rule. The saddle-point approximation is implemented by imposing the stationarity 
of the exponents with respect to the tensor fields $\sigma_{lm}^{'}$, $\sigma_{lm}^{''}$, $\rho_{lm}^{'}$ and  $\rho_{lm}^{''}$ and with respect to 
the molecular paths $y$ and $r$.  
In particular:
\begin{itemize}
\item
Imposing the stationarity with respect to the $r$ path leads to the equation 
\be
\label{SPE1}
M \ddot y &=& 2 M \gamma \dot y - 2 \frac{\partial}{\partial r}\left[ V\left(r + \frac{y}{2}\right) -  V\left(r - \frac{y}{2}\right)\right]\nonumber\\
&& - \sum_{l,m} \frac{\partial}{\partial r} \left[  f_{lm}\left(r + \frac{y}{2}\right)~\rho^{'}_{lm} - 
 f_{lm} \left(r - \frac{y}{2}\right)~\rho^{''}_{lm}\right] .\nonumber\\
\ee
\item
Imposing the stationarity with respect to the density tensor fields  $\rho_{lm}^{'}$ and $\rho_{lm}^{''}$ leads to the equations
\be \label{SPE2}
\sigma^{'}_{l m} &=&  f_{l m} \left[r+\frac{y}{2}\right],  \\
\sigma^{''}_{l m} &=& f_{l m} \left[r-\frac{y}{2}\right].
\ee
\item
Imposing the stationarity with respect to the conjugate fields  $\sigma_{lm}^{'}(t')$ and $\sigma_{lm}^{''}(t')$ leads to the equations 
\be
\label{SPE3}
&&\rho^{'}_{l m}(t') =\frac{1}{\sum_{k'} M_{k'}[\sigma^{''}]~Q_{k'}[\sigma^{'}]}~
 \sum_{k_f} ~M_{k_f}[\sigma^{''}] \int \D \phi^{'} \D \phi^{'*}  
\nonumber\\
&&\phi^{'}_{k_f}(t)~\phi_l^{*'}(t')\phi^{'}_m(t')~ \phi^{' *}_{k_0}(0)~e^{\frac{i}{\hbar} S_{CM}[\phi^{'},\phi^{'*},\sigma']},
\nonumber\\ \\
&&\rho^{''}_{l m}(t') = \frac{1}{\sum_{k'} M_{k'}[\sigma^{''}] Q_{k'}[\sigma^{'}]}  \sum_{k_f} Q_{k_f}[\sigma^{'}] \int \D \phi^{''} \D \phi^{''*}
\nonumber\\
&&  \phi^{'' *}_{k_f}(t)~\phi_l^{*''}(t')\phi^{''}_m(t')~\phi^{''}_{k_0}(0)~e^{-\frac{i}{\hbar} S_{CM}[\phi^{''},\phi^{''*},\sigma'']}.\nonumber\\
\ee
\end{itemize}

The set of saddle-point equations of motion (\ref{SPE1})-(\ref{SPE3}) are simultaneously satisfied if, for any $t'\in[0,t]$ one imposes
\be\label{SP1}
 y(t') &=& 0,\\
\label{SP2}
\sigma^{'}_{l m}(t') &=& \sigma^{''}_{l m}(t') = f_{l m}[r(t')],\\
\label{SP3}
 \rho^{'}_{lm}(t') &=& \rho^{''}_{l m}(t') \equiv \rho_{l m}(t').
\ee
Hence, backwards and forward evolution coincide, at the saddle-point level. 

Using Eq. (\ref{SPE3}) we note that $\rho_{lm}$ can be re-written as:
\be\label{evol1}
\rho_{l m}(t') &=& \langle \Psi_{[\sigma]} |~\hat a^\dagger_l \hat a_m ~| \Psi_{[\sigma]} \rangle\\
\ee
where the quantum state $|\Psi_{[\sigma]}\rangle$ is defined as:
\be\label{evolPsi}
|\Psi_{[\sigma]} \rangle &\equiv& T e^{-\frac{i}{\hbar}\int_0^{t'} d\tau~\hat H_{eff}[\sigma]}| k_0\rangle,
\ee
and the time-dependent Hamiltonian $\hat H_{eff}[\sigma]$  is defined as
 \be
\label{Heff}
 \hat H_{eff}[\sigma]=  \sum_{l,m}~ \sigma_{l m}(t)~  \hat a^\dagger_l \hat a_m.
  \ee 
  Hence, the field $\rho_{l m}$ is identified with the (reduced) density matrix, evaluated on the state  obtained by evolving for a time $t'$ 
the initial quantum state $|k_0\rangle$, according to the time-dependent Hamiltonian $\hat H_{eff}[\sigma]$.

\section{Quantum and Stochastic Equations of Motion}
\label{QMD}

In this section, we use the path integral representation of the conditional probability (\ref{Pcond}) and the saddle-point relationships (\ref{SP1})-(\ref{SP3}) to derive  a 
set of equations which describe the evolution of the quantum charge and the classical atomic nuclei, for a molecule in solution. 
 
Our strategy consists in estimating the path integral over the charge density field $\rho$ and its conjugate $\sigma$ field in the lowest-order saddle-point approximation developed in the  previous section. On the other hand, the integral over the molecular coordinates $y$ is evaluated at the one-loop level, i.e.  including
the effects of leading-order fluctuations around the saddle-point solution $y(t)=0$. This guarantees that the dynamics of the molecular coordinates $r$ is 
stochastic, 
even at the classical level. 

To implement this program, we impose that the density matrix field $\rho^{'}_{lm}=\rho^{''}_{lm}\equiv \rho_{lm}$ and its conjugate field $\sigma^{'}_{lm}=\sigma^{''}_{lm}\equiv \sigma_{lm}$ obey the saddle-point relationship (\ref{SP2}), (\ref{SP3}) and we focus on the remaining part of the path integral, which concerns the molecular  degrees of freedom,  $\D r$ and $\D y$,
\be\label{PIy}
\mathcal{P}_t(Q_f|Q_0;[\rho]) \equiv \int_{Q_0}^{Q_f}  \mathcal{D}r \int \mathcal{D}y~e^{\frac{i}{\hbar}\left(\mathcal{W}[r,y,  \rho]
+ i \hbar \Phi^{'}_{FV}[r,y]\right)} ,
 \ee
 where we have used Eq. (\ref{SP2}) to eliminate $\sigma_{l m}$.  

Retaining only the lowest orders in the expansion of  the $y(t')$ path around 
the saddle-point solution $y(t)=0$ the functional at the exponent reads
\be\label{LO}
&&\mathcal{W}[r,y,  \rho]+ i \hbar~ \Phi^{'}_{FV}[r,y] = \nonumber\\
&=& -\int_0^t dt' \left[~y(t') \cdot \left(M\ddot{r}(t')+ M\gamma \dot{r}(t') \right.\right. \nonumber\\
&& +\left. \left.  \frac{\partial}{\partial r}\mathcal{V}[r(t'),\rho(t')]  \right)  + \frac{i~ k_B T M \gamma}{\hbar} y^2(t^{'})\right],
\ee
where
\be
\label{mathcalV}
\mathcal{V}[r, \rho] = V(r ) + \sum_{l m}  \rho_{l m} f_{l m}(r) = V(r ) + \text{Tr}[ \hat f(r ) \hat \rho ]
\ee

Since we have kept only terms which are at most quadratic in  $y(t')$,  the path integral over  this variable  can be evaluated   analytically by completing the square. The resulting (unnormalized) expression
for the  path integral (\ref{PIy}) is 
\be\label{PIsto}
\mathcal{P}_t(Q_f|Q_0;[ \rho]) = \int_{Q_0}^{Q_f}  \mathcal{D}r 
~  e^{-\frac{\int_0^t dt^{'} 
\left\{ M\ddot{r}(t')+ M\gamma \dot{r}(t') + \frac{\partial}{\partial r}\mathcal{V}[r,\bar{\rho}] \right\}^2}{4 k_B T M\gamma}}.\nonumber\\
 \ee
We emphasize that the action in the exponent generalizes  the well-known Onsager-Machlup functional~\cite{OMfunc} which appears in the path integral representation of the 
classical Langevin dynamics (a brief review is given in the appendix). 

In macro-molecular systems in solution inertial effects are damped at a time scale $10^{-13}~s$, which much smaller than the time scale associated to local conformational changes.
If the acceleration term $ M\ddot r$ at the exponent is neglected 
the path integral (\ref{PIsto}) becomes
\be\label{PIsto2}
\mathcal{P}_t(Q_f|Q_0;[ \rho]) =\int_{Q_0}^{Q_f}  \mathcal{D}r 
~  e^{-\frac{M\gamma}{4 k_B T}  ~\int_0^t dt^{'} 
\left\{\dot{r}(t') + \frac{1}{M \gamma }\frac{\partial}{\partial r}\mathcal{V}[r,\bar{\rho}] \right\}^2}\nonumber\\
 \ee
 which is completely equivalent to the stochastic path integral defined in Eq. (\ref{PIOM}) for classical over-damped Langevin dynamics .
 
 Hence, we conclude that the evolution of the  system can be described by the following set of quantum and stochastic differential equations:
\be
\label{Result1}
\left\{
\begin{array}{rl}
&\frac{d}{dt} r_\alpha = - \frac{1}{M \gamma} \frac{\partial}{\partial r^\alpha}\left(V(r )+ \text{Tr}[ \hat \rho~\hat f(r )] \right)+ \eta_\alpha(t) \\
&\\
&\frac{d}{dt} \hat{\rho} = - \frac{i}{\hbar} [\hat f(r ),~ \hat \rho],\\
\end{array}
\right.
\ee
 and
 we have condensed the Eq.s (\ref{evol1}), (\ref{Heff}), (\ref{SP2}) and (\ref{SP3})  into a single  Van-Neumann Equation, where $[\hat \rho(t)]_{l m} = \rho_{l m}(t)$  
 and $[\hat f]_{l m}(r )= f_{l m}(r )$. $\eta^\alpha(t)$ is the usual white delta-correlated Gaussian noise of the Langevin dynamics,
 \be
 \langle \eta^\alpha(t) \cdot \eta^\beta(0) \rangle = 
6 \frac{k_B T}{ M \gamma}~ \delta^{\alpha \beta}~ \delta(t).
\ee 

The set of equations (\ref{Result1}) represents main result of this paper, as far as the formalism is concerned.
Some comment on these equations are in order.  First of all, we emphasize that the Langevin equation for the molecular coordinates  
contains the force term 
\be
 F_\alpha(Q) \equiv ~\text{Tr} \left[\hat \rho,  -\frac{\partial}{\partial r^\alpha} \hat f(r ) \right].
 \ee
 This term is of fully quantum origin, as it follows directly from the 
the Hamiltonian (\ref{Htot}). It expresses the influence of the charge distribution on the molecular motion: the atoms are driven towards configurations for which the 
energy of the charge is lower. This type of non-Coulombic charge-nuclei interaction is not included in the
 standard phenomenological approaches which have been  used to simulate the dynamics of molecular wires~\cite{DNAelsner, EDNA, Wallace}.

If the molecule contains charged atomic groups (e.g. like in DNA), one needs also to include their Coulombic interaction with the propagating quantum charge.  This can be rigorously done by extending the potential energy function $\hat V(Q)$ in the original Hamiltonain (\ref{HM}) to:
\be
\hat V(Q)\rightarrow && \hat V(Q; \{\hat a_l, \hat a_l^\dagger\}_{l=1, \ldots, N_s} )  \nonumber\\
&&\equiv V(Q) + \frac{1}{2}\sum_{\alpha=1}^{N_p} \sum_{l=1}^{N_s} 
\frac{q_\alpha ~e}{|r_l - r_\alpha|}~\hat a^\dagger_l \hat a_l,
\ee
where $e$ is the charge of the hole and $q_\alpha$ is the (partial) charge of the $\alpha$-th atom.  
At the level of the Langevin equation (\ref{Result1}), this coupling produces an additional Coulomb 
force term in the form:
\be
F^\alpha_c(r, \rho_{l m}) = - \sum_l \int d r_\alpha
\frac{e_\alpha e ~\rho_{l l}}{(r_l - r_\alpha)^2}~ \hat u_{l \alpha},
\ee
where $\hat u_{l \alpha} = (r_\alpha -r_l)/|r_\alpha -r_l|$.

The equations of motion (\ref{Result1}) can be integrated by simultaneously evolving the molecular positions 
and the initial quantum state, and then computing the corresponding density matrix field: 
\be
\label{Result2}
\left\{
\begin{array}{rl}
& r_\alpha(t+\Delta t) = r_\alpha(t) - \frac{\Delta t}{M \gamma} \frac{\partial}{\partial r^\alpha}\left(V[r(t)]
+ \text{Tr} [\hat \rho \hat f (r )]  \right) 
\\
&\qquad\qquad + \sqrt{2 \frac{k_B T\Delta t}{M \gamma}} \xi_\alpha(t) \\
&\\
&|\Psi(t+\Delta t) \rangle = e^{-\frac{i \Delta t }{\hbar}~ \hat H_{eff}[r(t)]}  | \Psi(t)\rangle,\\
&\\
& \rho_{l m}(t+ \Delta t) = \langle \Psi(t+\Delta t) | l \rangle \langle m | \Psi(t+\Delta t) \rangle,\\
\end{array}
\right.
\ee
where $\xi_a(t)$ is a stochastic variable sampled from a Gaussian distribution with zero average and unitary variance.

The solution of the set of equations (\ref{Result1}) through the algorithm (\ref{Result2}) yields
 the real-time evolution of the atomic nuclei and of the charge density, starting from an initial condition.
Like any approach based on the time integration of the equations of motion, this method is expected to be computationally inefficient to 
investigate rare thermally-activated reactions. 
The main difficulty arises from the fact that, on average,  the first reactive event occurs on a time scale 
which scales exponentially with the height of barrier.  Hence, very long trajectories have to be generated in order to investigate the dynamics of the reaction.

In order to  overcome these difficulties, the Dominant Reaction Pathways (DRP) method  was recently developed~\cite{DRP0, DRP1, DRP2, DRP3, DRP4}.
This approach uses the path integral representation of the classical Langevin dynamics to derive a variational principle which yields the most probable reaction pathways
connecting  given initial and final configurations. 
 To date, the DRP algorithm has been successfully applied to investigate a number of reactions which are extremely hard to simulate using standard MD algorithms, including  
conformational transitions of peptides from \emph{ab-initio} simulations \cite{folding2}, protein folding within realistic atomistic classical  models~\cite{folding1}, 
 as well as chemical reactions\cite{ab-initio} or particle-surface interactions~\cite{QDRP}.  Recently, the formalism has been extended to include quantum corrections to
 the motion of light nuclei\cite{QDRP}.

The Feynman-Vernon functional integral formulation of the dynamics developed in the previous sections 
offers the proper  framework to  extend the DRP method to the case in which the molecule contains quantum excitations.
The starting observation is 
that the solutions of the saddle-point equations (\ref{SP1})-(\ref{SP3})  and the molecular path minimizing the effective action 
in the exponent  of Eq. (\ref{PIsto})   yield the \emph{most probable} coupled evolution of the charge density and of the molecular coordinates
during transitions. 

Assuming that the initial and final molecular configurations and the initial position of the charge are held fixed, the saddle-point equations can be solved 
self-consistently by means of the following iterative algorithm: 
\begin{enumerate}
\item  A reaction pathway $r(t)$ connecting the initial configuration $Q_0$ and the final configuration $Q_f$ 
 is computed by minimizing the Onsager-Machlup action at the 
exponent of Eq. (\ref{PIsto2}), neglecting the coupling with the density matrix $\rho_{lm}(t')$. This can be efficiently done by using the algorithms described in 
Ref.s \cite{folding1, folding2}.
\item The deterministic evolution of the  density field $\rho_{l m}(t')$ is computed from Eq.s (\ref{evolPsi}) and (\ref{Heff}), where the tight-binding coefficients in 
$H_{eff}$ are evaluated along the path $r(t)$ obtained in the previous step.
\item The density matrix obtained in the previous step is plugged in Eq. (\ref{PIsto2}), and an improved estimate of the molecular reaction path $r(t)$ 
is obtained by numerical relaxation of the generalized Onsager-Machlup action~(\ref{PIsto2}).   
\end{enumerate}
The process is iterated until convergence.

\section{Charge Localization in a Collapsing Chain}
\label{application}

Having developed the appropriate formalism, we are now ready to analyze the quantum transport dynamics during the collapse of a homo-polymeric chain.

According to the standard scaling theory~\cite{Anderson}, the existence of an Anderson metal-insulator transition (MIT) in the thermodynamic limit and at zero 
temperature critically depends on the number of spatial dimensions. In particular, in the  one-dimensional Anderson model, an arbitrarily small
amount of disorder is sufficient to stop electric conduction. 
In three-dimensions, a mobility edge is developed and the system remains in the metallic regime for sufficiently small disorder. 
If such idealized conditions were good approximations for realistic molecular wires in solution, then one might argue that the collapse 
into a molten globule of a polymer may lead to an increase of the conductance.  

On the other hand, physical molecular wires in solution are complex mesoscopic systems subject to dynamical disorder and finite temperature effects. 
As a consequence, their conductance 
depends on a number of specific physical
 conditions, such as e.g. the chain length, heat-bath temperature, noise memory function,  and so on. 

Nevertheless,  this systems may still display different conducting properties,  depending if the polymer is in a coil configuration (where the transport dynamics is effectively 
one-dimensional) or in a compact phase ( where the charge can diffuse in three-dimensions). 

This discussion raises two main questions: does the dynamical disorder driven by the conformational 
fluctuations of the chain induce a strong charge localization already in the coil state? Is such a localization enhanced or reduced when the system 
reaches a compact configuration? 

To address these questions, we use the framework developed in the previous sections to 
simulate the charge migration in a simple coarse-grained  model for a polymeric collapse.  

\subsection{Definition of the Model} 
We consider a chain composed by $N_p= 36$ beads, each one representative of an individual monomer. These particles are linked by harmonic 
springs and can interact at distance
 through Van-der-Waals potential, which simulates the effective hydrophobic attraction and the steric repulsion:
\be
\label{VQmodel}
&&V(Q) =  \frac{1}{2} \sum_\alpha  k_s (|r_{\alpha+1}- r_\alpha |-a)^2\nonumber\\
&&+ \frac 12\sum_{\alpha \ne \beta} 4 \epsilon 
\left[\left(\frac{\sigma}{|r_{\alpha}-r_\beta|}\right)^{12} - 
~\left(\frac{\sigma}{|r_\alpha- r_\beta|}\right)^{6}\right]
\ee

The hole can propagate by hopping across $N_s$ sites, which are identified with the chain's monomers, hence $N_s=N_p$. The chain is not coupled to electrodes. 

In the present illustrative toy model, the couplings in the effective configuration-dependent tight-binding Hamiltonian (\ref{HMC})
are not evaluated microscopically from electronic structure calculations. Instead, their dependence on the monomer coordinatesis is defined phenomenologically as follows:
\be
\label{HMCmodel}
T_{i j} (Q) = T_0 ~e^{- \frac{|r_i - r_j|^2}{2 a_t^2}},\qquad
e_{i}(Q) = \epsilon_0. 
\ee
Note that the charge can hop between monomers which are specially close, even if they are far in the chain sequence.

The numerical value of the  constants entering Eq.s (\ref{VQmodel}) and (\ref{HMCmodel}) are listed in table~\ref{table}.  We consider a heat bath of 
 temperature $T= 300$~K and, while the mass and the friction coefficient are chosen in such a way that $ M \gamma= 2000$~amu~ps$^{-1}$. 
This set of parameters was chosen to ensure that the end-to-end propagation of the quantum charge across the wire and the collapse of the molecule
occur at comparable time scales.  

\begin{table}[t!]
\begin{center}
\caption{Parameter of the Hamiltonian for the toy-model molecular  wire}
\label{table}
\begin{tabular}{|c |c| c| c| }
\hline
$k_s$  & $a$ & $\epsilon$  & $\sigma$\\
$[$kJ mol$^{-1}$ nm$^{-2}$$]$ & [nm]  & [kJ mol$^{-1}]$  & [nm]\\
\hline
1000 & 0.38 & 4 & 0.4  \\
\hline
\hline
$T_0$ & $\epsilon_0$& $a_T$ &  \\
$[$kJ mol$^{-1}$$]$ & [kJ mol$^{-1}$] & [nm] & \\
\hline
2 & - 200 & 0.3 &\\ 
\hline
\end{tabular}
\end{center}
\end{table}%

\subsection{Quantum Transport and Conformational Dynamics}

The potential energy function of this model is characterized by a low degree of frustration. As a result, the collapse of the chain from an initial stretched coil 
configuration to a compact globular state occurs  in just few ns. Such a time interval 
 can be simulated directly by integrating the equation of motion~(\ref{Result1}).

\begin{figure}[bt]
\begin{center}
\includegraphics[width=9cm]{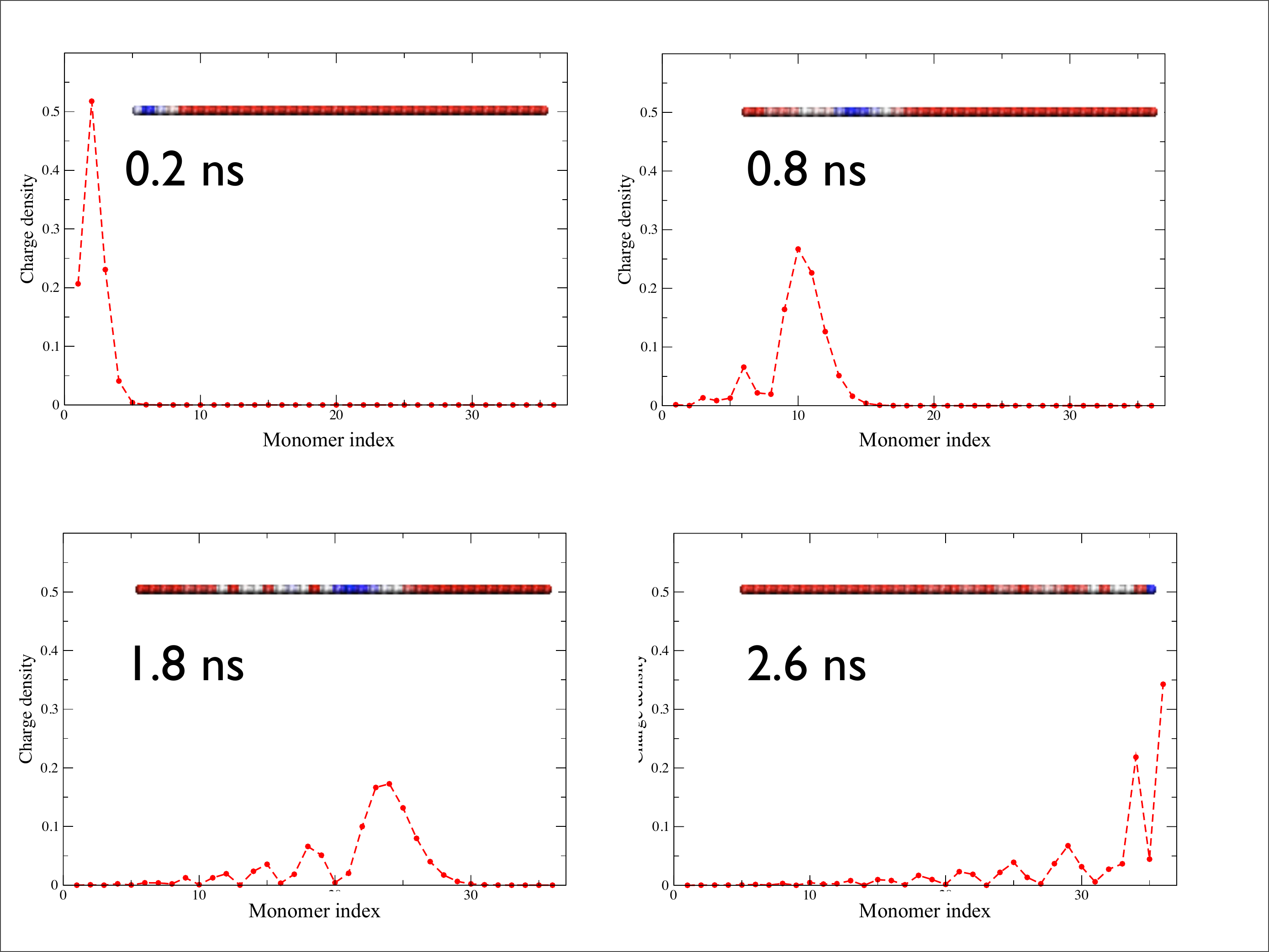}
\caption{Time evolution of the charge density distribution in a frozen fully stretched polymer. The pictures in the upper-right corner represent the polymer conformation, the color of the 
backbone is proportional to the charge density.}
\label{evolution_static}
\end{center}
\end{figure}

Let us consider a  polymer which is initially prepared in the fully stretched configuration shown in
 Fig. \ref{evolution_static}, with a quantum charge localized at the first monomer $|\Psi(t=0)\rangle = | 1\rangle$. 
 Then,  the evolution of the system is predicted by means of the algorithm (\ref{Result2}). 

The elementary time-step in the integration of the Langevin equation for this simple coarse-grained model can be chosen to be $\Delta t_{MD}= 10~$fs,
while the  time-step in the quantum evolution of the charge wave-function has to be  chosen much smaller,  e.g.   $\Delta t_{Q} = 1/10 \Delta t_{MD}$. 
With such a choice of parameters, simulating few ns of dynamics takes only a few minutes on a standard laptop computer and the normalization of the charge
  wave-function remains constant within $1 \%$ accuracy over the entire simulated time  interval.

Let us begin by discussing the charge migration dynamics in a static wire, i.e. assuming that the molecule remains 'frozen' in the initial fully stretched configuration.
This configuration corresponds to a complete absence of disorder. 
In Fig. \ref{evolution_static} we report the calculated evolution of the charge density in few ns of dynamics.
We note that the propagation of the charge occurs through an attenuating traveling density wave, which reaches the end of the wire in about 3 ns. 
The secondary peaks  are due to the fact that, at any instant, the charge has finite probability not to perform the transition. In the long time limit, the primary peak looses intensity, which signals that the charge is delocalized, with a consequent increase of the entropy. 

\begin{figure}[tbp]
\begin{center}
\includegraphics[width=9cm]{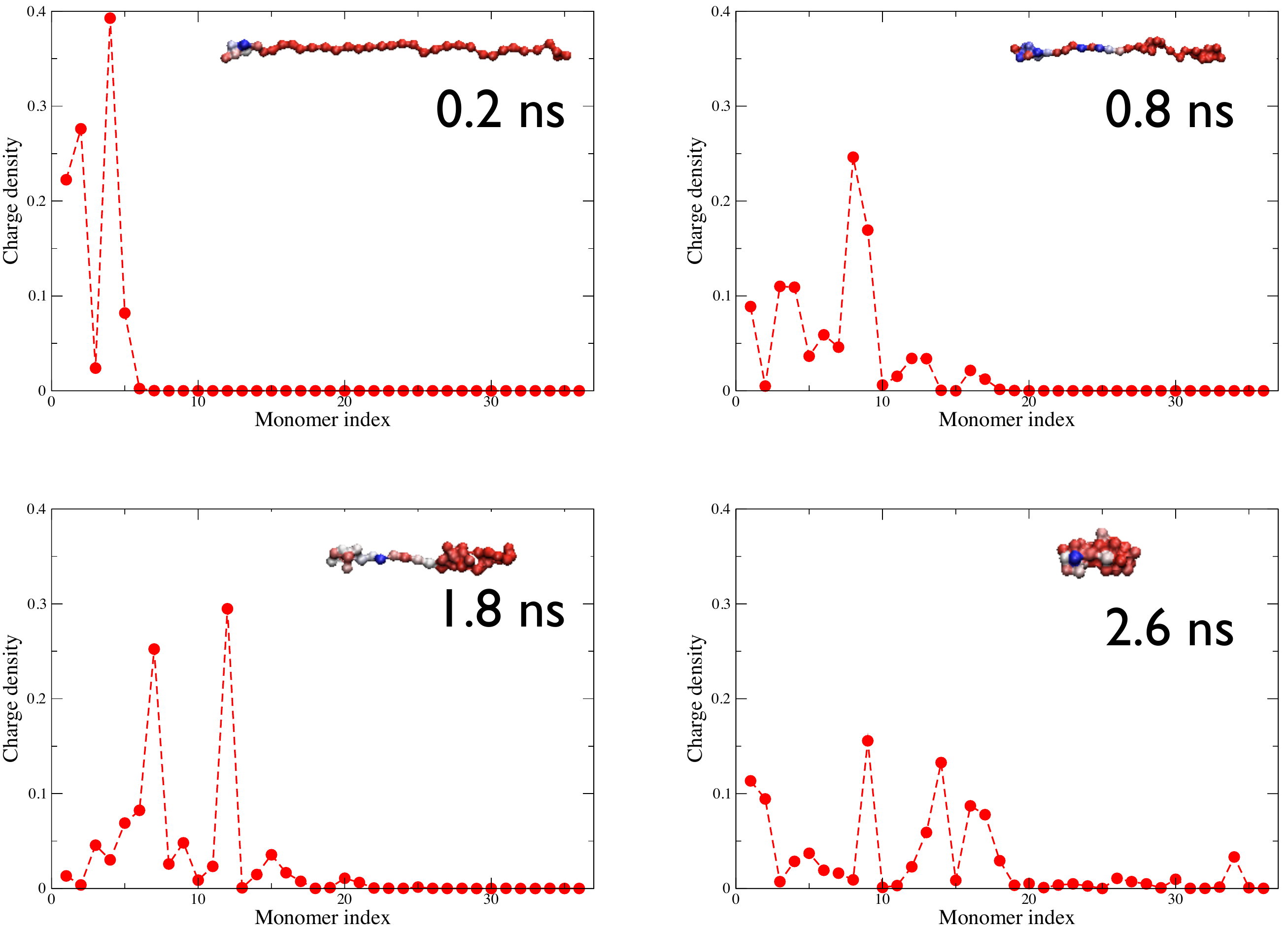}
\caption{Time evolution of the charge density distribution in collapsing polymer. The pictures in the upper-right corner represent the polymer conformation, the color of the 
backbone is proportional to the charge density.}
\label{evolution_dynamics}
\end{center}
\end{figure}

Fig. \ref{evolution_dynamics} shows that the quantum transport dynamics is radically changed if the polymer is allowed to move under the effect of the internal forces and of the fluctuations induced by the solvent, which introduce dynamical disorder in the tight-binding Hamiltonian.  We can clearly see that the migration of the charge is strongly hindered 
and that the initial density peak is completely dispersed after about one ns and hardly propagates into the second half of the chain. 
Hence, the charge transmission properties of the system are strongly affected by the configurational dynamics.


These results raise the question if the disorder-driven charge localization takes place already in the coil state, or only when the chain reaches the compact state. 
This question can be addressed in the context of random matrix theory~\cite{Mehta, Mirlin}.
This theory predicts that the level spacing distribution of a real orthogonal Hamiltonian with delocalized eigenstates should obey the so-called Wigner-Dyson distribution\footnote{Note that, as usual, this distribution is normalized in such a way to enforce a unit average level spacing. }, 
\be
P_{WD}(s) = \frac{\pi s}{2} e^{- \frac{\pi~s^2}{4}}.
\ee
In particular, we note  that $P_{WD}(s)$  vanishes for $s\to0$. This feature  is called level repulsion and is a characteristic signature of delocalized eigenstates (i.e. of the
 metallic regime).

By contrast, the localization of the wave-functions reduces the overlap between adjacent modes, hence suppressing the level repulsion. 
In particular,  in the extreme case in which there is no overlap between neighboring eigenvectors, the level-spacing distribution follows 
a Poisson distribution:
\be
P_P(s) = e^{-s}.
\ee

We have computed the level spacing distribution by simulating  48 independent  3 ns-long evolutions starting from the same stretched coil configuration, with the 
charge initially localized at the left terminus of the chain. The level spacing distribution  at different instants during the transition was then obtained by
explicitally diagonalizing the quantum Hamiltonian (\ref{HMC}). 

\begin{figure}[htbp]
\begin{center}
\includegraphics[width=8cm]{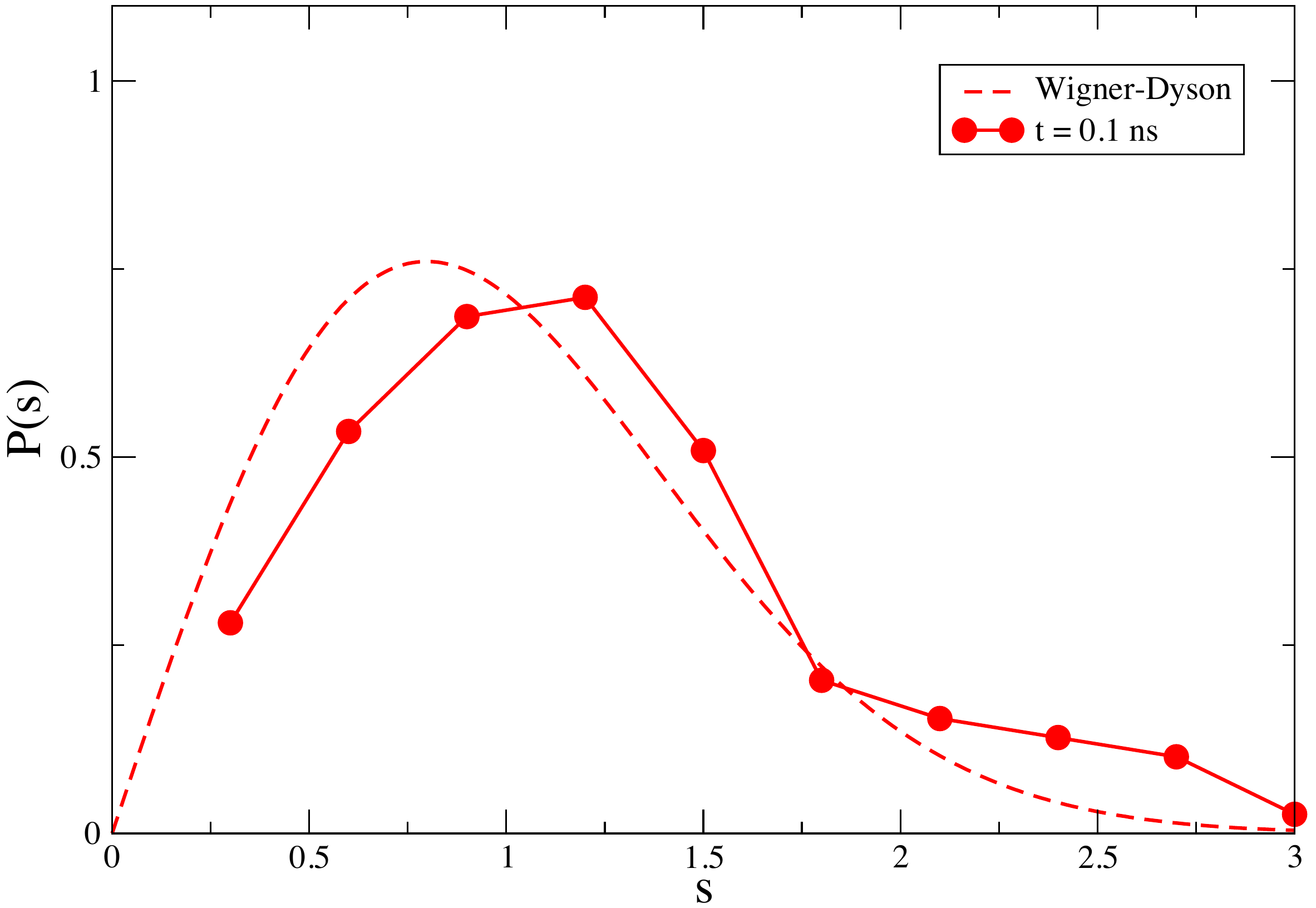}
\includegraphics[width=8cm]{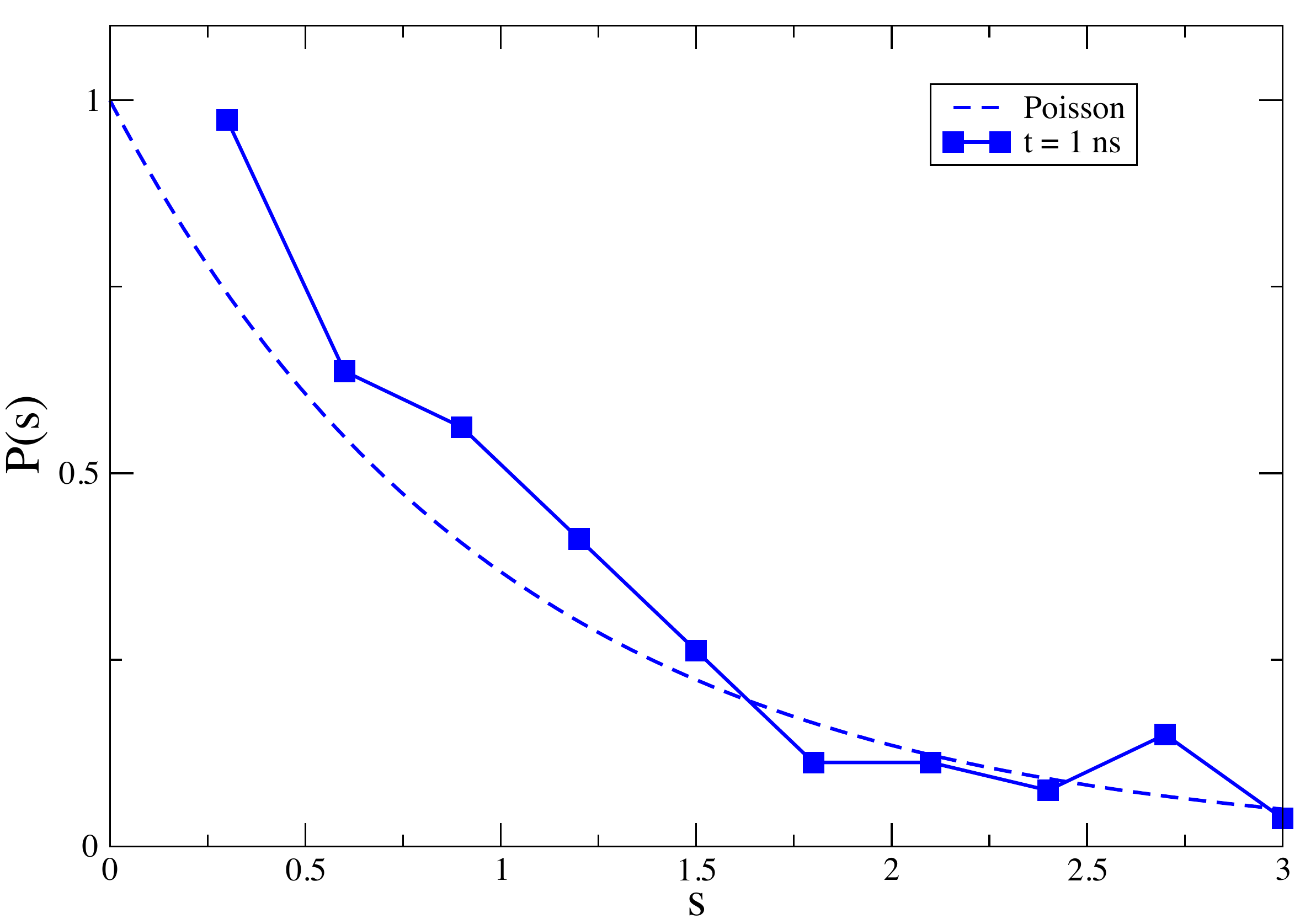}
\caption{Level spacing distribution for the tight-binding Hamiltonian, evaluated after  after 0.1 ns (upper panel) and after 1 ns (lower panel)..}
\label{RMTfig}
\end{center}
\end{figure}

The results are summarized in Fig. \ref{RMTfig}. In the early time of the collapse the spectrum is well explained by a Wigner-Dyson statistics, hence the system displays good conducting properties
even in the presence of disorder. On the other hand,
after about 1 ns, revel repellence has completely disappeared and the spectrum is well described by a Poisson distribution, hence the system behaves like an insulator.

Note that the observed behavior contrasts with what would be  expected from scaling arguments, implying that the meso-scopic character of this system and thermal effects play an important role.

 The recent developments in single-molecule and ensemble manipulation techniques may allow to look for this effect.
For example, using a conducting atomic force microscopy it is  in principle possible  to simultaneously measure   the conductivity  \cite{cAFMDNA, cAFMproteins} and the end-point distance
\cite{AFMendpoint1,AFMendpoint2} of single polymers.
Alternatively, by simultaneously applying  laser-induced temperature jumps~\cite{LJ} and electron pulse 
radiolysis~\cite{charging} on an ensemble of neutral collapsed polymers it is possible to drive them into 
non-equilibrium swollen and electrically charged states. 
Then, the evolution of the conductivity in these systems during the subsequent thermal relaxation process may be monitored using time-resolved microwave conductivity techniques~\cite{inorganic4}. 

\section{Conclusions}
\label{conclusions}

In this paper, we have presented a formalism to quantitatively investigate  the dynamics of quantum excitations inside macromolecules which move under the 
effect of the intramolecular forces and of the collisions with the solvent molecules. The resulting
set of coupled stochastic and quantum equations of motion (\ref{Result1}) contains a non-Coulombic force term which is not present in the approaches where 
equations of motion are postulated phenomenologically.  Such a term has a quantum origin and explicitly couples 
the density matrix of the quantum particle to the molecular coordinates. The numerical simulations performed in a simple coarse-grained model have shown that this
force can give significant contribution, of the same order of standard (i.e. Van-der-Waals) non-bonded interactions. 

We have also used the path integral formalism for open quantum systems to derive an algorithm which yields the most probable pathways in the evolution of this
 system, from a given initial to a given final molecular conformation.   This formalism may provide a computationally efficient method to investigate the non-equilibrium quantum transport dynamics during rare thermally-activated conformational transitions.
  
We  have applied this formalism to study the migration  of a quantum charge, during the coil-globule transition of  a  polymer.
These calculations have shown that the dynamics of the quantum charge is strongly quenched during the collapse. By analyzing the statics of the eigenvalues spectrum, 
we have concluded that the disorder-driven increase of localization sets in only when the system reaches a compact conformation.   

The formalism developed in this work does not only concern molecular quantum wires, but can be applied  to all molecular systems in solution which can support the propagation of quantum excitations. In particular, in the future it would be interesting to use it to investigate the transfer of neutral atomic excitations between amino-acids in peptide chains. This would provide 
a solid theoretical ground to bridge the gap between molecular simulations and experimentally 
observable circular dichroism and F\"orster Resonance Energy Transfer spectra in protein folding.  

\appendix
\section{Path Integral representation of the classical dynamics of a molecule in a thermal bath}

The main goal of  this work is to derive a path-integral based  framework to study the evolution of quantum excitations in conformationally evolving molecular systems.
 
For comparison purposes, it is instructive to review the corresponding path integral formulation of the classical  Langevin dynamics which is often used to study the conformational dynamics in the absence of quantum excitations. The so-called 
over-damped Langevin equation reads\cite{Schwabl}:
\be
\label{Langevin}
\dot q_\alpha = - \frac{1}{M \gamma} \frac{\partial }{\partial q_\alpha} V(Q) + \eta(t),\qquad (\alpha=1,\ldots, N_p)
\ee  
where $\gamma$ is the friction coefficient, $V(Q)$ is the potential energy function entering in Eq. (\ref{HM}) and $\eta(t)$ is delta-correlated Gaussian noise, satisfying the fluctuation-dissipation relationship:
\be
\langle \eta^\alpha(t')\cdot \eta^\beta(t) \rangle =   \frac{6 k_B T}{M \gamma}~\delta^{\alpha  \beta}~ \delta(t-t')\qquad (\alpha, \beta = 1, \ldots, N_p).
\ee
Note that in the original Langevin Eq. there is a mass term $M \ddot q$. However, for macro-molecular systems this term is damped at a time scale $10^{-13}~s$, which much smaller than the time scale associated to local conformational changes.

The stochastic differential Eq. (\ref{Langevin}) generates a time-dependent probability distribution $P(Q,t)$ which obeys the well-known Smoluchowski Eq.:
\be
\label{FP}
\frac{\partial}{\partial t} P(Q,t) = \frac{k_B T}{ M\gamma} \nabla \left[ \nabla P(Q,t) + \frac{1}{k_B T}\nabla V(Q) P (Q,t)\right].
\ee

By performing the formal substitution 
\be
P(Q,t)= e^{-\frac{1}{2 k_B T} V(Q)}~\Psi(Q,t),
\ee
the Smoluchowski Eq. (\ref{FP}) can be recast in the form of an imaginary time  Schr\"odinger Eq.:
\be
 -\frac{\partial}{\partial t} \Psi(Q,t) = \hat{H}_{eff}~\Psi(Q,t),
\label{SEA}
\ee
where
\be
\hat{H}_{eff}~=~- \frac{k_B T}{M \gamma} \hat{\nabla^2} +\hat {V}_{eff}(Q),
\label{HeffL}
\ee
  is an effective Hamiltonian operator and  
\be
V_{eff}(Q)= \frac{1}{4 k_B T M \gamma }\left( (\nabla V(Q))^2- 2 k_B T \nabla^2 V(Q)\right).
\label{Veff}
\ee

The conditional probability $P(Q_f, t| Q_0)$  to find the system at the configuration $Q_f$ at time $t$, provided it was prepared in the configuration $Q_0$ at time $t=0$ is the Green's function of the  Smoluchowski Eq., and  can be related to the imaginary time   propagator of the effective "quantum" Hamiltonian (\ref{HeffL}):
\be
P_t(Q_f|Q_0) &=&  e^{-\frac{1}{ 2 k_B T} (V(Q)-V(Q_0))} ~   \langle Q_f | e^{- t H_{eff} }| Q_0\rangle.
\label{K}
\ee
Using such a connection, it is immediate to obtain an expression of  the conditional probability (\ref{K}) in the form of a Feynman path integral
\be
\label{PI}
P_t(Q_f|Q_0) = e^{-\frac{V(Q_f)-V(Q_0)}{2 k_B T} } ~\int_{Q_0}^{Q_f} \mathcal{D} r~e^{-S_{eff}[r]},
\ee
where 
\be\label{Seff}
S_{eff}[r] =\int_{0}^t dt' ~\left(M \gamma~\frac{~\dot{r}^2}{4 k_B T}  + V_{eff}[r]~\right)
\ee
is called the effective action. 

The conditional probability $P_t(Q_f|Q_0)$ is sometimes written also in the following equivalent form:
\be
\label{PIOM}
P_t(Q_f|Q_0) = \int_{Q_0}^{Q_f} \mathcal{D} r~e^{-S_{OM}[r]},\nonumber\\
\ee
where $S_{OM}[r]$ is the so-called the Onsager-Machlup functional,
\be\label{SOM}
S_{OM}[r] =\int_{0}^t dt' ~\frac{M \gamma}{4 k_B T}\left(\dot r + \frac{1}{M \gamma} \nabla V(r)\right)^2,
\ee 
Proving the equivalence between the expressions (\ref{PI}) and (\ref{PIOM}) is not straightforward, since it involves elements of stochastic calculus ~\cite{Adib, MDlowk}. 

\acknowledgments
PF is a member of the Interdisciplinary Laboratory for Computational Science (LISC), a joint venture of Trento University and FBK foundation. 

PF acknowledges an important discussion with M. Sega, S. Kantorovich and A. Arnold. 
We also thank our colleagues F. Pederiva, G. Garberoglio, S. Taioli and L. Pitaevskii for useful comments.

\end{document}